# Polygonal micro-whirlpools induced in ferrofluids

Revised 4/23/2015  15:11:00

Marcin Bacia[1], Weronika Lamperska[1], Jan Masajada[1], Sławomir Drobczyński[1], Maciej Marc[2]

[1]*Wrocław University of Technology, Department of Optics and Photonics*
*Wybrzeże Stanisława Wyspiańskiego 27, 50-370 Wrocław, Poland*
*e-mail address: marcin.bacia@pwr.edu.pl*
[2]*Institute of Physics, University of Zielona Góra, Szafrana 4a, 65-069 Zielona Góra, Poland*

We report on the observation of the polygonal whirlpools in the thin layer of ferrofluid under illumination with a laser beam carrying optical vortex and in the presence of a vertical magnetic field. This kind of structures have attracted attention after discovering a hexagonal storm in Saturn's atmosphere. Our polygonal whirlpools were created in a closed system (no free surfaces) in micro scale (whirlpool diameter<20μm) by the use of holographic optical tweezers. The polygonal shape was changed by varying the magnetic field strength or value of the optical vortex topological charge.

The Saturn's hexagonal storm was first discovered during the Voyager mission nearly 30 year ago [1] and recently confirmed by the Cassini-Huygens mission [2]. This hexagonal-shaped cloud pattern surrounds Saturn's North Pole and rotates around the planet's axis. Its movement remains surprisingly stable, which makes the storm different from the hurricanes on Earth. Various theories were presented in order to describe its origin and motion [3-4].

In the laboratory conditions the polygonal whirlpools were generated using a cylindrical container with water [5-6]. The flow was forced from the bottom by rotating the mechanical system. When the rotation rate became sufficiently large, the axial symmetry of the free surface was broken and large time dependent deformations appeared. The surface of the rotating fluid was formed into regular polygons with 3-5 corners.

In this paper we describe a creation of polygonal whirlpools at microscopic scale (<20μm of diameter) in a thin droplet of ferrofluid (~12μm of thickness). The ferrofluid has no free surface (it is closed between two microscopic slides). Ferrofluids are colloidal suspensions of magnetic particles having average size of about 10nm dispersed in liquid (organic one in our case) [7-8]. Magnetic particles are coated with a surfactant to inhibit their aggregation and sedimentation. Ferrofluids have both liquid and magnetic properties, which are useful for various applications. The light action on the ferrofluids was described in [7, 9-11] (paper [7] contains a review on optical properties of ferrofluids). The review on colloids research with optical tweezers (but not holographic ones) can be found in [12]. In our experiments we used commercial and homemade ferrofluids. The observed effects in both cases were the same.

Interestingly enough, we need three factors for creating microscopic polygonal whirlpools in ferrofluid. These factors are: heat transfer, angular momentum transfer to the magnetic particles, and vertical magnetic field. All these factors are present in the Saturn's atmosphere. Certainly, this coincidence does not mean that the effects observed in the ferrofluids are a miniature copy of the hexagonal storm in the Saturn's atmosphere. However, the hexagonal storm has attracted wide attention, so we think that all potential clues are worth disseminating. Besides we believe that our observations are also interesting outside the Saturn's storm context.

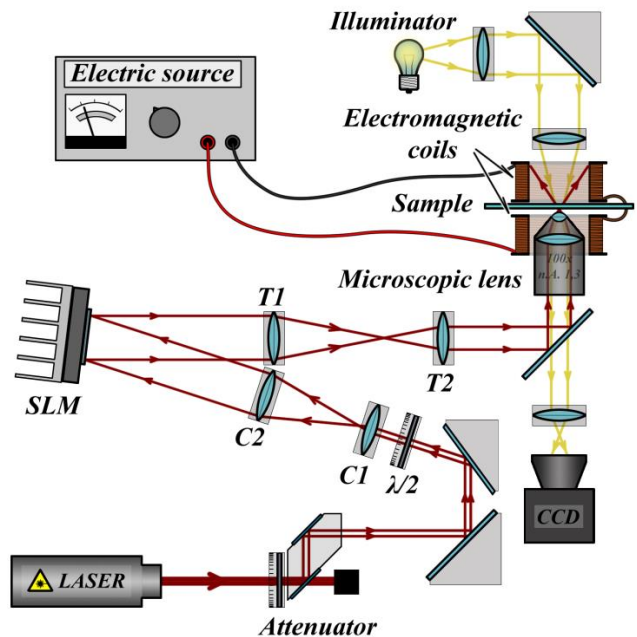

FIG. 1. (Color online) Scheme of the holographic optical tweezers with two electromagnetic coils.

To create polygonal whirlpools in ferrofluids we used holographic optical tweezers [13,14]. They allows creating tens of optical traps within the area of tightly focused laser beam. The scheme of the optical setup used in our experiments is shown in Fig. 1. It is based on reverse biological microscope Olympus IX71. The optical trap was generated at the sample plane by the use of the spatial light modulator (SLM, Holoeye-Pluto). The SLM was illuminated by the laser beam (1064nm) and then the beam was focused by immersion microscope objective Olympus UPlanFL N (100x) with NA=1.3. In the sample plane the magnetic field was generated by the use of two electromagnetic coils. With the holographic optical tweezers different kinds of traps (light, dark [15] or Bessel [16]) can be generated simultaneously. Traps can be also moved independently in *x-y-z* direction.

Small particles, both metallic and dielectric, can be trapped by such a focused light beam. In the case of metallic particles this effect was described in [17-18]. In our experiment we work with dark traps, i.e. focused laser beams carrying optical vortices. The vortex beam carries non-zero angular momentum which is related to its helical wavefront shape [19]. The angular momentum is proportional to the parameter called topological charge (here denoted by letter *m*). Due to the transfer of the angular momentum (from the beam to the particles), small particles can spin [20] and rotate around the central part of the trap [21]. It is hard to observe this circular motion directly due to small size of nanoparticles. To visualize this effect we did an experiment with a magnetic nanoparticles (average diameter 2nm) but synthesized in mesoporous silica (MCM-41) using the method of co-precipitation of $Fe^{2+}$ and $Fe^{3+}$ chlorides (in molar ratio 1:2) in an alkaline medium [22]. Thus, the large number of nanoparticles was closed in quasi-spherical silica structure of diameter ~1μm and the effect of interaction between light and silicon spheres with magnetic particles could be directly observed.

The silica spheres dispersion in water and oil was prepared and closed between two microscopic slides. The layer thickness was ~12μm. The sample was illuminated with a dark trap (*m*=20). The light intensity distribution within such a trap is shown in Fig. 2a - it has a characteristic dark centre surrounded by the bright ring. Figures 2b-e show that under laser illumination mesoporous silica particles were trapped along the bright ring. The whole chain of particles rotated around the beam center. We observed a rotation of the small gap in the particle chain (indicated by white arrowhead in Fig. 2b-e). In this experiment the magnetic field was off. When the optical vortex was oriented reversely (the topological charge of the vortex is of the opposite sign, *m*=-20) the silica spheres with the magnetic nanoparticles rotated in opposite direction. The rotation speed depends on both: the beam intensity and the value of the optical vortex topological charge. We also observed that the pure MCM-41 spheres (free of magnetic particles) dispersed in water (or oil) were briefly trapped (unstable trapping), went around about quarter of the loop and escaped from the trap. That means that stable trapping depends on the presence of magnetic nanoparticles. The results were the same for both water and oil dispersion, but in oil the particles rotation was slower.

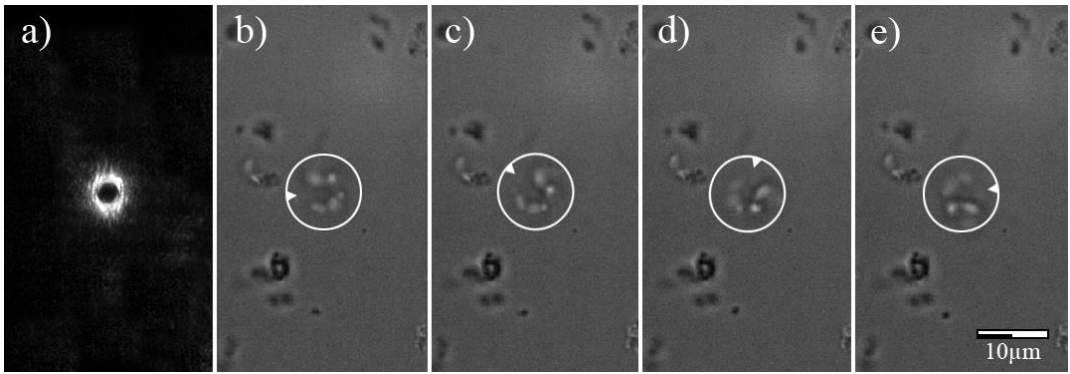

FIG. 2. a) The shape of the vortex beam at the sample plane; b-e) Chain of mesoporous silica spheres with magnetic nanoparticles gathered around the vortex beam center (water dispersion). There is a small gap in this chain indicated by white arrowhead. Pictures were taken at 0,24s intervals.

In the next step the thin ferrofluid (~12μm) layer was illuminated with the dark trap and the vertical magnetic field was applied. The polygonal whirlpool with bright arms emerging from the polygon corners was created and the whole pattern rotates (Fig. 3). The rotation occurred only when a dark trap was applied. The number of arms depended on vortex charge, light intensity and magnetic field strength. Figure 4 shows a dependence on optical vortex charge (the magnetic field is constant) and Figure 5, a dependence on magnetic field strength (*m*=40).

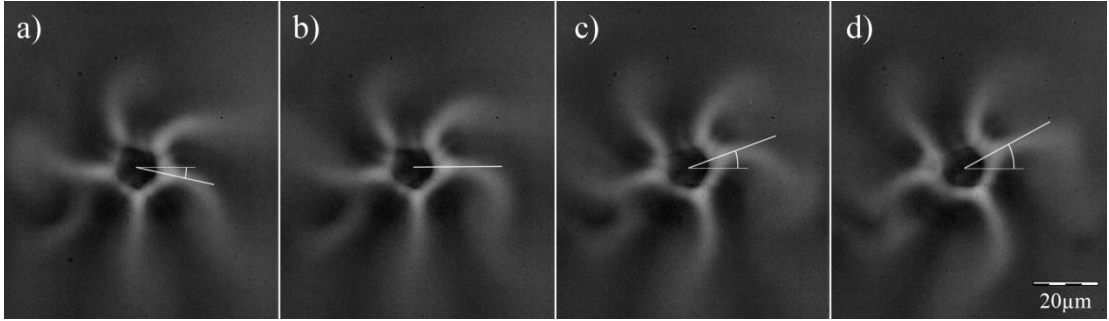

FIG. 3. Vertical magnetic field and dark trapping beam forms a pentagonal form which rotates. Images were taken at 0,7s intervals. This phenomenon can be also seen on Video 1.

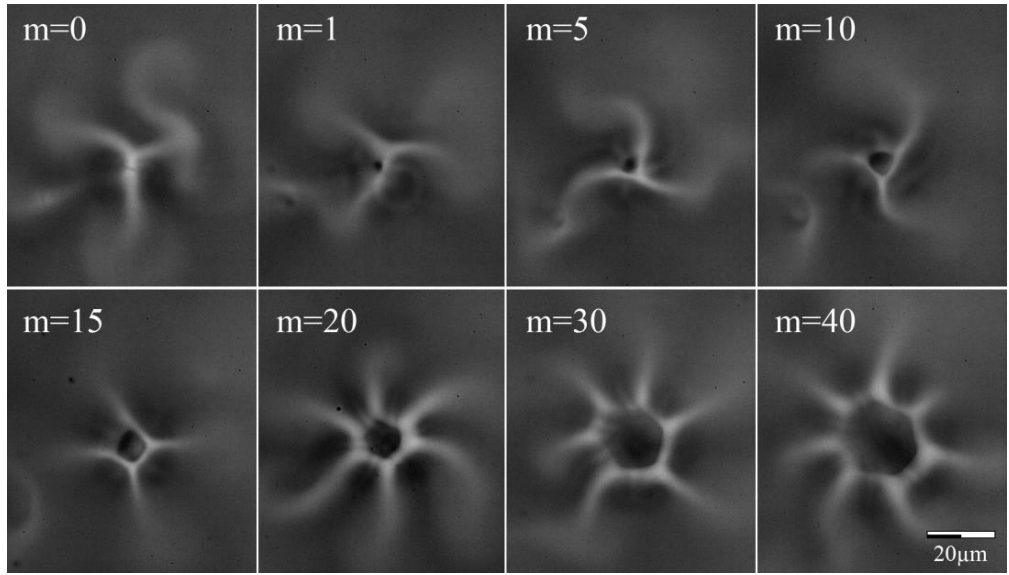

FIG. 4. The polygonal geometry depends on the topological charge of the trapping beam. If m=0 there is no polygon and the structure does not rotate.

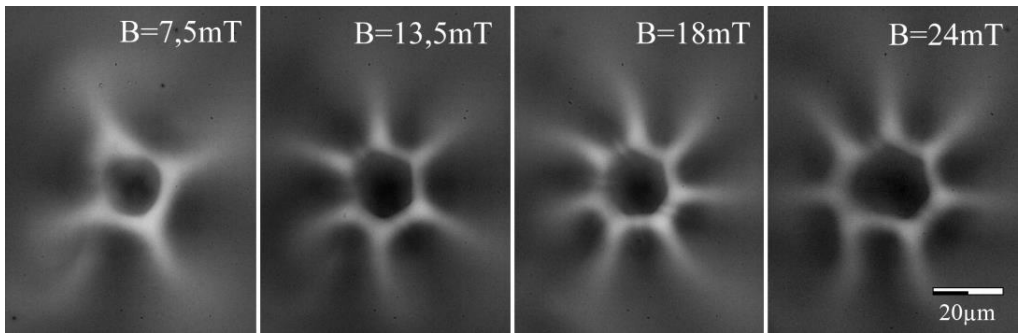

FIG. 5. The polygonal geometry depends on the magnetic field strength. The vortex beam topological charge is $m=40$.

However, this dependence is not exact. For given vortex charge, light intensity and strength of magnetic field, the number of arms varies by one showing that the whole phenomenon is highly sensitive to sample preparation, sample age and environmental conditions.

In the absence of the dark optical trap, the magnetic field is not enough to cause polygonal whirlpools (Fig. 4: m=0). The pattern has bright arms emerging from its center, number of which depends on magnetic field strength and light intensity. However the whole pattern does not rotate.

It is also worth noticing that the rotation of white arms occurred only in the presence of the vertical magnetic field. The horizontally oriented magnetic field stretches the bright pattern into a single straight line that remains motionless (Fig. 6). This effect has already been reported in [23].

Polygonal patterns in water at macroscopic level were studied in paper [4]. It was shown that small whirlpools are located by each side of the polygon. The careful observation of our polygonal structures revealed similar motion. However, it is hard to see it directly. We added various micro-particles to the ferrofluid to make the effect well visible. Best results were obtained with small (1μm) particles of carborundum. The presence of small whirlpools by each polygon side is revealed by circular movement of the carborundum particles (Fig. 7).

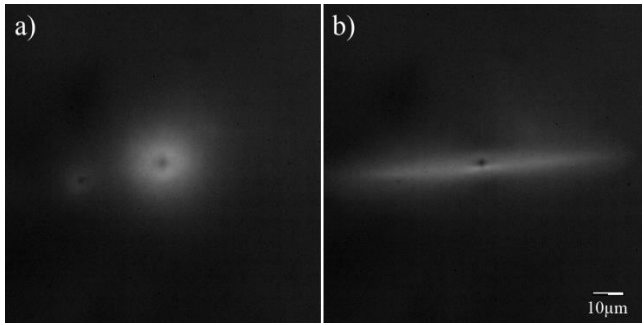

FIG. 6. No polygon forms are created with horizontally oriented magnetic field. a) the sample illuminated without magnetic field and with horizontally magnetic field (b). In both cases the dark trap illuminated the sample.

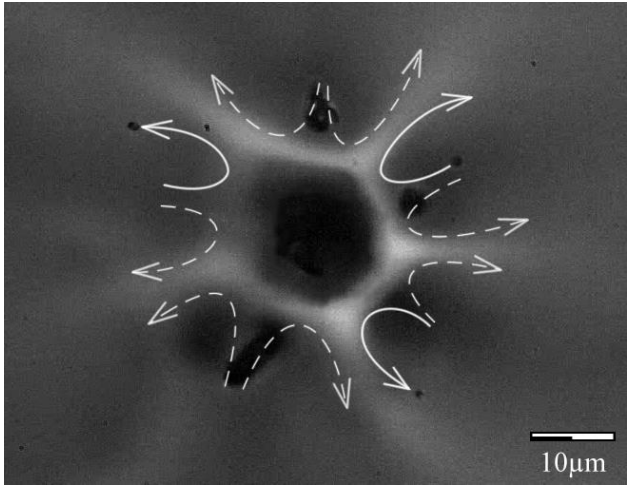

FIG. 7. Carborundum particles start to circulate close to the polygon sides. The solid white lines show the trajectory of three particles captured at the moment the picture was taken. The dotted white lines show the trajectory of particles captured in other time intervals.

Performing the experiments we noticed that when the ferrofluid layer was too thin (below 6 μm) the whirlpools became irregular most probably due to adhesion (between liquid and slides). Too thick layer also blocked the effects reported in this paper.

The other factor changing the character of the observed patterns is the energy density of the beam. The heat transfer causes the particles move bottom – up. When the energy density of the beam is too high the gas bubble is created in the beam centre and the magnetic particles are pushed out. The observed pattern looks different (Fig. 8) due to very strong spinning of nanoparticles around the gas bubble [24]. We created several independent traps within observation plane. Each trap generated separate bubble which interacted in a complicated way. They are joined with bright lines along which the magnetic particle aggregates are transported (Fig.8). Similar effects were observed in $p$-nitroaniline dissolved in 1,4-dioxane [25].

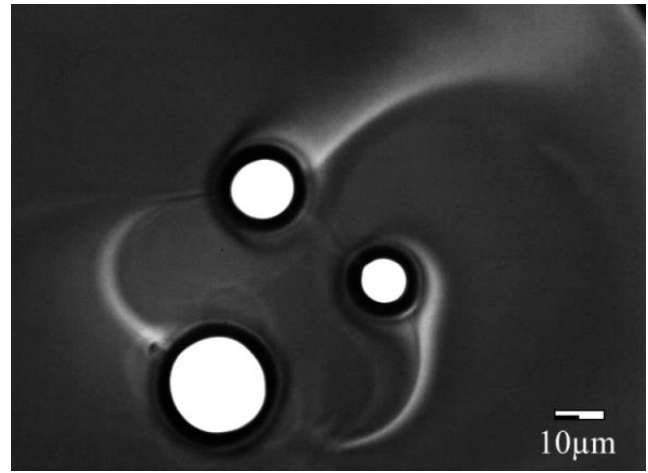

FIG. 8. Three gas bubbles created by three optical traps of high energy density. The whirlpools associated with gas bubbles interact which each other and exchange particles along the bright lines.

In conclusions: The polygonal whirlpools at microscopic scale can be generated in the thin layer of ferrofluid illuminated with laser beam carrying an optical vortex and in the presence of vertical magnetic field. The ferrofluid droplet is closed between two microscopic slides, so the free surface is not necessary for this phenomenon to be observed. The number of the polygon sides depends on the optical vortex charge, light intensity, and the strength of the magnetic field. The rotation speed of the whole pattern depends on optical vortex charge and light intensity. However, the laser light intensity is limited by the effect of gas bubble creation. When the gas bubble is created no polygonal forms can be observed.